\documentclass[fleqn,10pt]{wlscirep}
\usepackage[utf8]{inputenc}
\usepackage[T1]{fontenc}
\usepackage{ulem}

\definecolor{revgreen}{rgb}{0.0,0.55,0.0}
\newcommand{\rev}[1]{\textcolor{revgreen}{#1}}

\title{An AI-Enabled Agent-Based Simulation Platform for Studying COVID-19 Pandemic}

\author[1,+]{Anik Burman}
\author[2,+]{Sayak Chatterjee}
\author[3,*]{Pramit Ghosh}
\author[4,*]{Indranil Mukhopadhyay}
\affil[1]{Johns Hopkins University, Department of Biostatistics, Baltimore, MD 21205, USA}
\affil[2]{University of Pennsylvania, Department of Statistics and Data Science, Philadelphia, PA 19104, USA}
\affil[3]{ICMR-National Institute for Research in Bacterial Infections, Department of Epidemiology, Kolkata,700055, India}
\affil[4]{University of Nebraska-Lincoln, Department of Statistics, Lincoln, 68683, USA}

\affil[*]{imukhopadhyay2@unl.edu, ghosh.00@gmail.com}

\affil[+]{these authors contributed equally to this work}


\begin{abstract}
Understanding the dynamics of an outbreak like that of COVID-19 is important in designing effective control measures. This study aims to develop an agent based model that compares changes in infection progression by manipulating different parameters in a synthetic population. Model input includes demographic characteristics like age, sex, working status of each individual along with other factors influencing disease dynamics. Depending on number of epicentres of infection, location of primary cases, sensitivity, proportion of asymptomatic and frequency or duration of lockdown, our simulator tracks every individual, and hence infection progression through community over time. In a closed community of $10000$ people, it is seen that without any lockdown, number of cases peak around 6th week and wanes off around 15th week. If primary case is located inside dense population cluster like slums, cases peak early and wane off slowly. With introduction of lockdown, cases peak at slower rate. If sensitivity of identifying infection decreases, cases and deaths increase. Number of cases declines with increase in proportion of asymptomatic cases. The model is robust and provides reproducible estimates with realistic parameter values. It also guides in identifying measures to control outbreak in a community. It is flexible in accommodating different parameters like infectivity period, yield of testing, socio-economic strata, daily travel, awareness level, population density, social distancing, lockdown etc. and can be tailored to study other infections with similar transmission dynamics. We make this model available as an open, interactive web application that lets users without any programming background design scenarios and observe outbreak dynamics in real time. Beyond forecasting, we position the simulator as a reusable \textit{in-silico} environment, or digital twin, that can serve as a synthetic-data generator and a decision-support platform for artificial-intelligence-driven optimization of intervention policies.
\end{abstract}

\begin{document}

\flushbottom
\maketitle
\thispagestyle{empty}

\section{Introduction}

The COVID-19 pandemic left a lasting impression on those who survived. It transcends the personal sphere and marks an inflection point in biomedical research worldwide. Everyone was searching for realistic forecasts and effective control measures during those chaotic days. The efforts did play a pivotal role in shaping interventions across the globe, and trust in models to guide policy was reinforced.

Merit of a model lies in its utility. In this work, we have developed a model to understand the disease dynamics in the backdrop of population composition, interaction patterns and available interventions within a community.  Research and innovation in the domains of diagnostics, devices, drugs and vaccines are progressing at a rapid pace. Research advancement will definitely influence transmission as well as consequence of the infection. Hence, for a model to remain relevant,  it should have a generic framework,  flexible enough to accommodate relevant changes synchronous with scientific advancement. 
In a real world scenario, benefit of scientific advancement seldom percolate down to the end user uniformly; a mix of bio-social factors not only determine the risk of disease acquisition, but also influence accessibility or availability of health interventions. The variable risks of contracting disease and availing benefits of interventions provide modelers an opportunity to explore optimized solutions in a given context. This model allows comparison between competing mix of interventions and aids decision making.

In retrospect, the global and local responses to the COVID-19 pandemic can be considered as a cauldron of natural experiments; the outcomes of interventions during the period can alter our a-priori assumption to aid in prevention and management of future pandemics. COVID-19 is a viral disease, primarily transmitted through respiratory droplets and aerosols. However, the route is not specific to COVID-19. The pan-India integrated surveillance captured information from 34,260 cases of Severe Acute Respiratory Illness (SARI) and Influenza-Like Illnesses (ILI) between July 2021 to October 2023\cite{potdar}; Contribution from the influenza virus and COVID-19 was comparable. There are several other pathogens which closely mimic COVID-19, and a study compared five such organisms viz. Influenza virus, Respiratory Syncital Virus, Rhinovirus, Human coronavirus and Adenovirus, through a modeling approach \cite{spencer}. The parameter values, timing, prevalence, and proportional contributions differ substantially. These results showed the importance of correct identification of responsible pathogen. The virus often changes the subtypes with variable transmission risks and virulence. Hence technology-enabled surveillance efforts need to be strengthened during the course of the outbreak also so that we can keep track of changes within the pathogen. In this backdrop we conceptualize a flexible AI-oriented framework that includes inherent stochasticity and dynamics of the disease phenomenon to observe the effect of interventions on the progression of an outbreak of respiratory infection taking COVID-19 as an example.

Commonly used models using a set of differential equations like SIR \cite{huang2020epidemic,leonardi2020self}, describe epidemiological state transitions from susceptible to infectious to recovered compartments. Mathematical epidemiology is flooded with extensions and sophistication \cite{mwalili2020seir,ghosh2020increased,yang2020modified,das2020critical} sprouting out from SIR models that include subdivision of exposed class, hospitalization, host-pathogen interaction, and so on. Undoubtedly, incorporation of other compartments would make the disease dynamics model more realistic; on the other hand, it involves too many parameters that need to be known and/or estimated a-priori. These models are used for prediction of compartment sizes, impact and severity of the disease etc \cite{roberts2015nine}. Hence accurate knowledge about these parameters are essential as the performance of such models entirely depends on parameter values. Although they provide good estimates, the initial sizes of different compartments, especially the susceptible class, strongly influence the dynamics of the infection progress. Standard estimation procedure based on available data often fails to account for time dependent parameters \cite{roberts2015nine}.

The interplay of various factors like age, socio-economic strata, and occupation of a person along with availability, accessibility and efficiency of testing and treatment would eventually determine the propagation and impact of the disease in our community and country \cite{lakshmi2020factors}. This dynamic and complex nature of the system and society involves many variables that not only shows non-linear behavior but are also time dependent \cite{das2020critical}. Inherent stochasticity and spatio-temporal behavior of the factors involving the disease dynamics might lead to lack of accuracy of a deterministic model in tracing the spread of disease over time. Keeping in mind the basic structure of these popular and useful models like SIR, SEIR etc, agent-based modeling (ABM) becomes another tool \cite{cuevas2020agent,hoertel2020stochastic,epstein2012agent} to capture the disease dynamics in the presence of several constraints on the parameters. 

Agent-based modeling is a simulation-based computational paradigm deeply rooted in artificial intelligence. In a complex adaptive network of multi-agent systems, dynamics of autonomous agents is governed by local decision rules manifesting into population-level behavior. It has the potential to capture spatio-temporal evolution of the system involving socio-economic variables, interactions with each other, and non-linear behavior under complex conditions that is sometimes very hard to describe mathematically. ABM can track the chain of infection depending on population characteristics \cite{supriya2013abm,liu2015,supriya2015abm,espana2018,guclu2016}. It can be an efficient tool in identifying the best possible interventions in a given situation \cite{wong2011automatic}. It runs through an environment where learning algorithms can be trained, evaluated and produced realistic output. Apart from being a forecasting device, ABM has the potential to act as a controllable yet flexible virtual laboratory on which AI methods like reinforcement learning for policy optimization, machine learning for reliable inference and simulation-based calibration, can be built.

The temporal, spatial and demographic distribution of COVID-19 disease reflects its inherent heterogeneity like any other disease, depending on its genetic nature, host characteristics and environmental correlates. Inter and intra-country variation is common as recently published article argues that urban agglomeration can play a crucial role in fueling and sustaining the pandemic across the globe along with overcrowding, occupation etc.\cite{desai2020urban} ABM can closely mimic this real-life situation despite several such complexities.

We aim to develop a common digital platform to facilitate simulation over a range of parameters. It is developed for the study of disease dynamics of influenza like illness, such as COVID-19. However, it would also help in understanding the propagation of any infectious disease in a specific community by a pathogen of similar nature spread by direct person-to-person transmission without any intermediary vectors. Our model is a general one as it involves parameters of an SEIR model in a broad sense, population demography, economic status, age, consciousness to follow some policies that are expected to contain the disease spread, treatment protocol and other factors and their interactions. Effects of the factors considered in our model might be of immediate nature or can unfold gradually over time as well. The platform itself is delivered as a freely accessible, no-code web application which is a central contribution of this work. It lowers the barrier to scenario analysis for public-health practitioners, educators and researchers, and provides immense flexibility for the AI-enabled extensions outlined above.

\section{Method}

We develop a generic agent-based model (ABM) for COVID-19 that may be applicable to any geographical location and socio-economic structure. This AI-enabled framework is flexible, with all model parameters specified as inputs that can be tailored to the characteristics of a particular community, locality, or country.

Input included both population characteristics and other factors influencing disease dynamics. Susceptible population was stratified based on age, sex, working status, type of work etc. These are all potential determinants for natural history of the disease. Risk of infection and transmission potential were thought to vary with education, profession, scope of social interaction, stage of infection etc. Agents or individuals with varying attributes behave differently and simulation over time reflect the propagation pattern.

\begin{figure}
    \centering
    \includegraphics[width=0.5\linewidth]{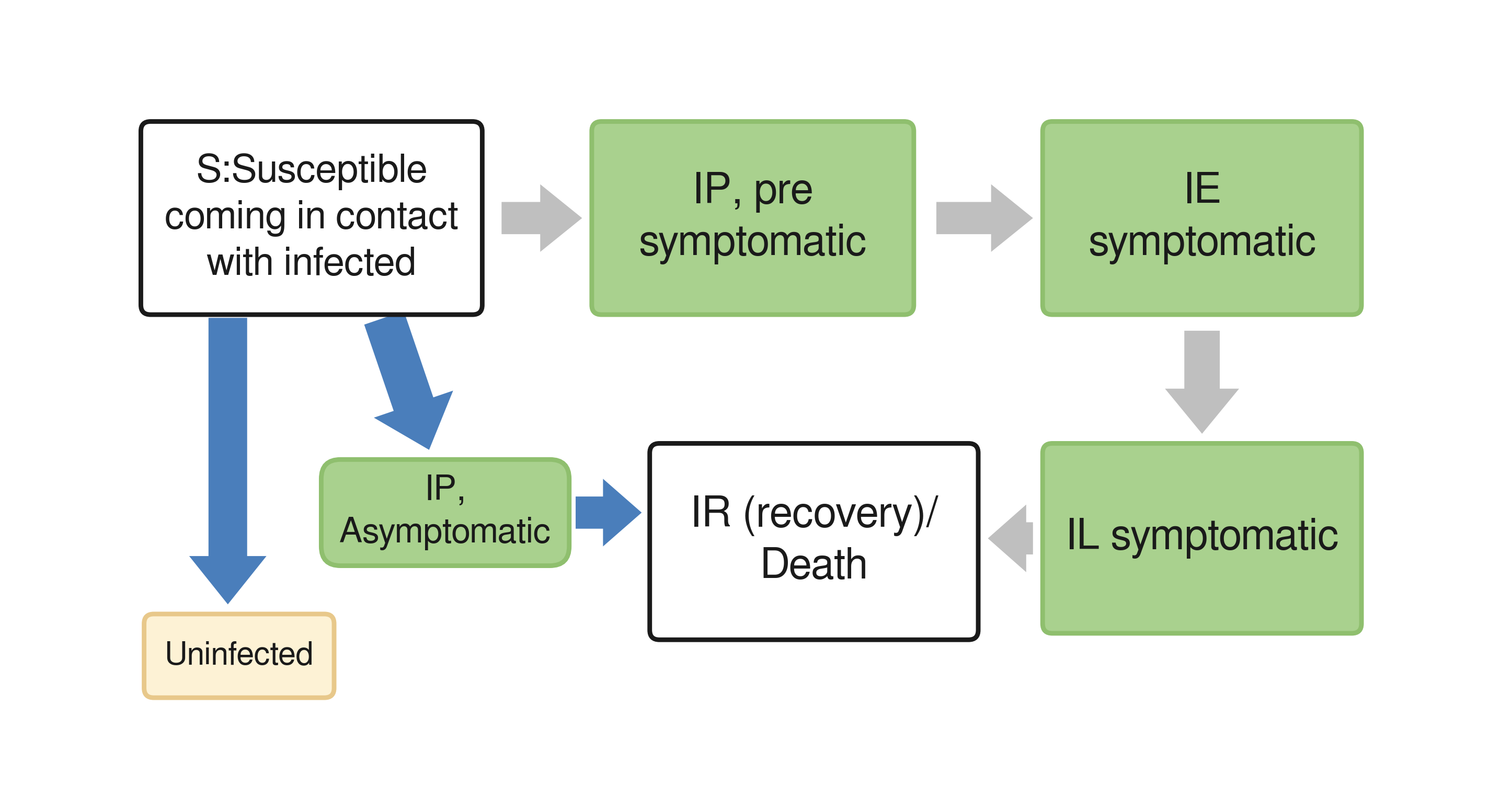}
    \caption{Transmission dynamics of the model.}
    \label{fig:model}
\end{figure}

Figure \ref{fig:model} describes various states through which a person moves during the course of the disease. Infected individuals can either remain asymptomatic throughout the 4-week period or manifest disease at some point of time albeit with varying potential. For symptomatic individuals, shedding of virus starts in pre-symptomatic phase, increases till first 3-4 days of manifestation and slowly wanes off in next 2-3 weeks. Force of infection, can be considered as an interplay of (a) rate of contacts among individuals, (b) probability of a contact being with an infectious host, and (c) the probability of that contact giving rise to an infection.

Our simulator tracks every individual, known as `agent' over each day in the entire time period since the inception of infection till it subsides significantly. Movement or mobility of an agent depends on age, working status, disease status, comorbidity, intention to follow lockdown etc. We also consider varying percentage of asymptomatic individuals that play a major role in disease spread. Once an agent dies, it is removed from the population and simulation continues without that agent. Comorbidity, as a function of age and other factors, is plugged into the risk probability that a person at a given stage will live one unit of time.

Considering all possible combinations of parameters as depicted in Table \ref{tab:param}, we have $72$ simulation models. Each model is run $1000$ times over a period of $150$ days in $20 \times 20$ sq unit area with $10000$ population, to see the average behavior for each scenario. Depending on the population density, we can vary its size and/or population size so as to mimic the actual population density. Local population density itself is a function of multiple socio-economic parameters. This also dictates the pace of disease transmission in any community. Social distancing norms were promoted to increase physical distancing so that virus could not get transmitted to a susceptible host \cite{world2020advice}. We assumed that once recovered, a person will not be re-infected during the course of the simulation. The movement from one stage to other may be considered as a continuum through which a person or agent traverses in unidirectional pattern, once infected. It is not necessary that all agents in the population will be infected; certain percentage may remain uninfected in the community throughout the course of simulation.

The model starts with an assumption that in a virgin community when the first infected person entered, all others were susceptible.  We restrict ourselves only to closed population assuming that no one will leave the population during the time period of simulation. Since spread of disease depends on population density, we assume 2- and 3- cluster models for population and varying number of epicentres and their locations. Within a cluster, population density is very high compared to other parts. Epicentres may lie within and/or outside a cluster. We also consider different periods of initiation and duration of lockdown and varying degree of precision for testing results. Our simulator incorporates the information on age and occupation distribution of the population in the region.

Initially at time 0, we assign age, economic and work status randomly to the population in each group. Sizes of different disease statuses (Figure \ref{fig:model}) are $n_i, \,\,i=1,\ldots,6$ such that $\sum_{i=1}^6 n_i=N$, where $N$ denotes the total population size. Since the time unit is one day and time frame is relatively small, we do not assume recruitment and natural death in this population. Moreover, we assume that the population is closed within the region considered as there are some restrictions in the form of lockdown, quarantine etc. Hence in our model inter-region mobility is not allowed. Our main interest is to see the disease transmission and dynamics within a micro environment region.

In the model, the user can select number of clusters and their locations within the rectangle at the start of simulation. Such clusters contain $20\%$ of the total population and the rest $80\%$ population is distributed uniformly over remaining region. Within clusters, the population is distributed uniformly. Note that it is easy to implement the cluster size in term of geographical region and assign population accordingly. However, we can safely assume the same population density within clusters which is much more than the population density in rest of the region. This assumption is valid because we know that the population in slum is much denser compared to other areas. This idea of introducing population density is based on the fact that disease is supposed to be transmitted faster in congested areas.

We assume that the immunity is low for early and old ages and highest in the middle age groups. We introduce this concept in terms of probability that a person is immune to the disease increases first, reaches a peak and then decreases with respect to age. Conditional probability of the test being positive given that the agent is actually infected, is taken as the sensitivity of the system in identifying an infected person.

Mobility and hence transmissibility along with other influencing factors also depend on gender and related factors. We assign the gender to an agent using a Bernoulli random variable with probability $p_G$ as the probability of begin male. The value of $p_G$ can be found for the region; usually it is close to 0.5.

Let $S_{kl}(x,y)$ and $I_{kl}(x_0,y_0)$ be a susceptible and an infected agent with location coordinates $(x,y)$ and $(x_0,y_0)$ respectively. Then $S_{kl}(x,y)$ belongs to the neighbourhood of $I_{kl}(x_0,y_0)$ if the euclidean distance between the two is less than a preassigned distance $l_0$ (say), {\it i.e.}
$$S_{kl}(x,y) \in \mathcal{N}(I_{kl}\big(x_0,y_0)\big) \text{ if } \sqrt{(x-x_0)^2 + (y-y_0)^2} \leq l_0 \text{ for all } (x,y)\in \mathcal{R} \text{ and } (x_0,y_0) \in \mathcal{P}$$
where $\mathcal{P}$ is the entire path covered by $I_{kl}$ due to its mobility at time $t$ and $\mathcal{R}$ is the entire region. Note that the circular region with centre at $(x_0,y_0)$ and radius $l_0$ must entirely be contained in $\mathcal{R}$; otherwise an agent may be outside the geographical region considered violating our model assumption.

Our simulator is implemented in R using the Shiny framework. It is deployed as a publicly accessible, browser-based application, so that the entire scenario-design-to-results workflow can be carried out interactively without any programming. In this sense, the tool functions as an \textit{in-silico} laboratory, or decision-support platform, rather than a one-off analysis script. We report an average curve with a band as given by the standard deviation based on 1000 repetitions of each model over the period of $150$ days.

The model has the scope of incorporating infection timeline, working and economic status, gender, age along with number and duration of lockdown periods. This makes it very flexible and adaptable to any geographical location with varying population density and socio-economic status.

\begin{table}
\centering
\caption{Parameters used in the ABM simulations.}
\label{tab:param}
\begin{tabular}{llccc}
\toprule
\textbf{No. of} &
\textbf{Location} &
\textbf{Detection} &
\textbf{Asymptomatic} &
\textbf{No. of} \\
\textbf{epicentres} &
 &
\textbf{sensitivity} &
\textbf{fraction} &
\textbf{lockdowns} \\
\midrule
2 & Within cluster  & 0.75 & 0.10 & 0 \\
3 & Outside cluster & 0.85 & 0.25 & 2 \\
  &                 & 0.95 & 0.50 &   \\
\bottomrule
\end{tabular}
\end{table}

\section{Results}

We present a few case studies that serve a dual purpose. They illustrate how the platform is used and, at the same time, provide a face-validity check that the simulator reproduces qualitatively well-established features of respiratory-epidemic dynamics. By altering parameters like location of primary case, sensitivity of diagnosing cases, proportion of asymptomatic cases or period and number of lockdowns, we observe changes in shape and spread of the outbreak.

\begin{figure}[ht]
    \centering
    \includegraphics[width=0.9\linewidth]{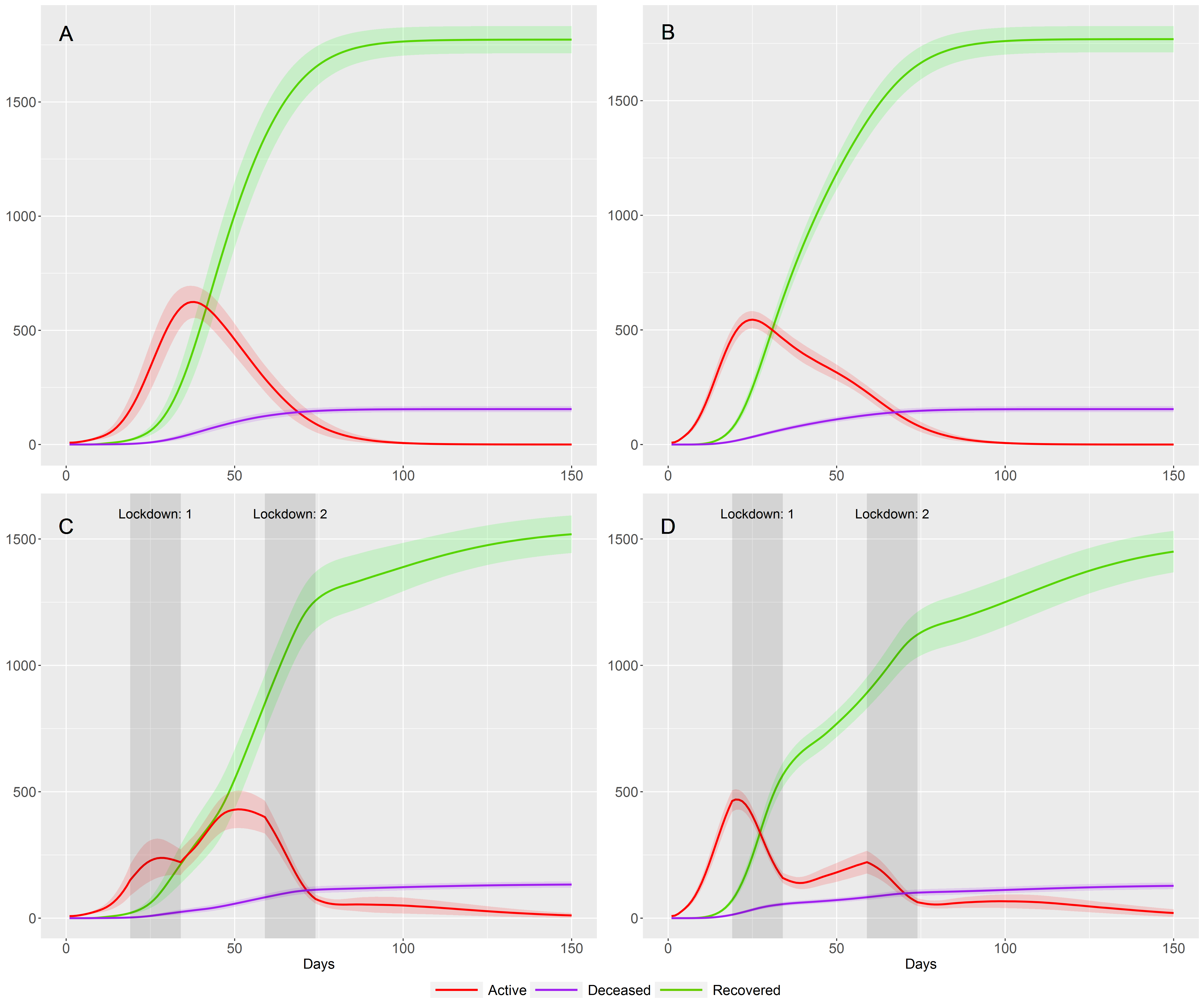}
    \caption{Daily active cases and cumulative deceased and recovered cases over time.}
    \label{fig:cases-1}
\end{figure}

Once the disease gets introduced in a community, the epicentre can either be in a cluster, i.e.  densely populated slums, or outside the clusters, i.e. in an airport terminal away from a residential area. We consider a situation where there are two new cases outside cluster in a virgin community (Figure \ref{fig:cases-1}A). We find cases peak around $35^{\text{th}}$ day with cases as high as 600 in a single day; curve gets flattened by $100^{\text{th}}$ day. Cumulative recovered cases were about 1750 and approximately 150 persons died from the disease.

Then we implement lockdown for 20 days twice, starting at $20^{\text{th}}$ day and $60^{\text{th}}$ day. This drastically altered shape of the curve (Figure \ref{fig:cases-1}C). At the end of the first lockdown, there is a little dip in active cases, followed immediately by a rise. Active cases peak around 50 days and maximum number of cases is around 450. With the introduction of second phase of lockdown, cases decrease fast, and eventually die out by $100^{\text{th}}$ day of outbreak.

A comparative study of these two graphs provides a good reflection of how effective lockdown is, even in a heterogenous population with variable chance of following the rules, and how it can deter and diminish the rise in active cases. This definitely buys time for community to prepare for the coming onslaught and reduce the chance of acute hospital bed crisis.

When cases start from within clusters, then without lockdown, outbreak pattern remains similar to the earlier (Figure \ref{fig:cases-1}B); albeit active cases peak earlier around 25 days and wanes off more slowly,  and eventually flattens around $100^{\text{th}}$ day. Recovered cases are also similar around 1750 (Figure \ref{fig:cases-1}B). Once we introduce lockdown, cases start to decrease from the beginning of lockdown. Curve rises a bit once first lockdown is lifted; but with introduction of second phase of lockdown, it decreases quickly (Figure \ref{fig:cases-1}D). There is another important difference seen. The curve does not touch x-axis even at $150^{\text{th}}$ day of simulation. Active cases are very few, but cases are still occurring. Cumulative number of recovered cases are around 1450 and graph of recovered cases is yet to plateau.

This indicates, that if started in densely populated areas, the outbreak will have a long-drawn course even in a small community of 10000 population even in spite of multiple lockdowns.

\begin{figure}[ht]
    \centering
    \includegraphics[width=0.9\linewidth]{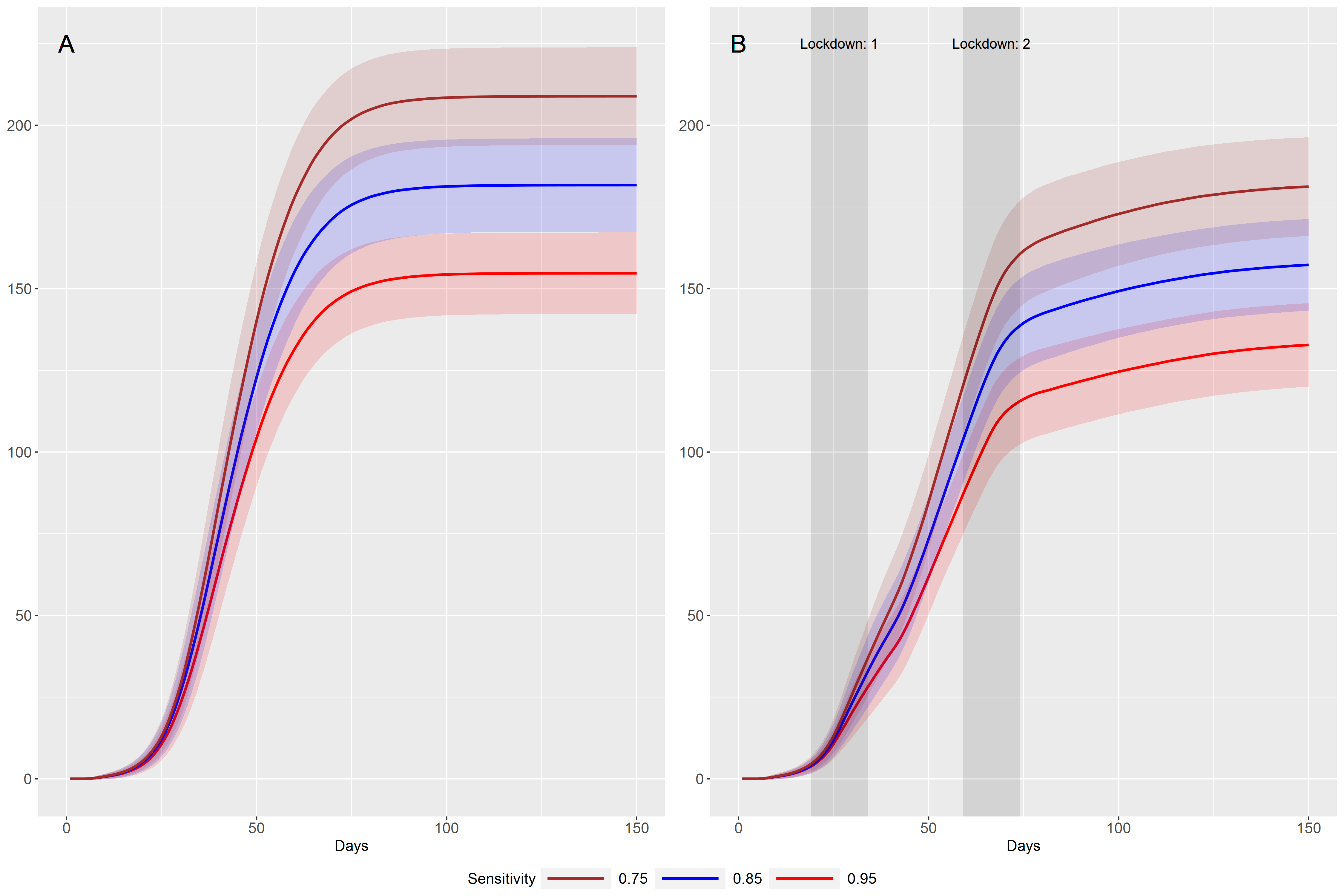}
    \caption{Cumulative number of deaths over time for different sensitivity values with and without lockdown.}
    \label{fig:cases-cum}
\end{figure}

In all these situations, we consider asymptomatic cases as $10\%$ of total cases and chance of detecting infected cases as $95\%$. If we lower the chance of identifying infected persons, there is an increase in the number of deaths from disease in the community (Figure \ref{fig:cases-cum}A) without any lockdown. With $95\%$ chance, cumulative death is around 125, which increases to 175 with $85\%$ and around 225 with $75\%$ chance. So, detecting cases through laboratory testing by escalating availability and accessibility of good diagnostic kits is critical to reduce mortality.

We consider another situation with $10\%$ asymptomatic infection and all primary cases are located outside population cluster. We implement two lockdowns and changed the sensitivity (Figure \ref{fig:cases-cum}B). Sensitivity has the most prominent effect on the number of deaths, which increases with decreasing sensitivity. The numbers are comparatively less with two lockdowns vis-a-vis none, keeping other parameters the same.

\begin{figure}[ht]
    \centering
    \includegraphics[width=0.9\linewidth]{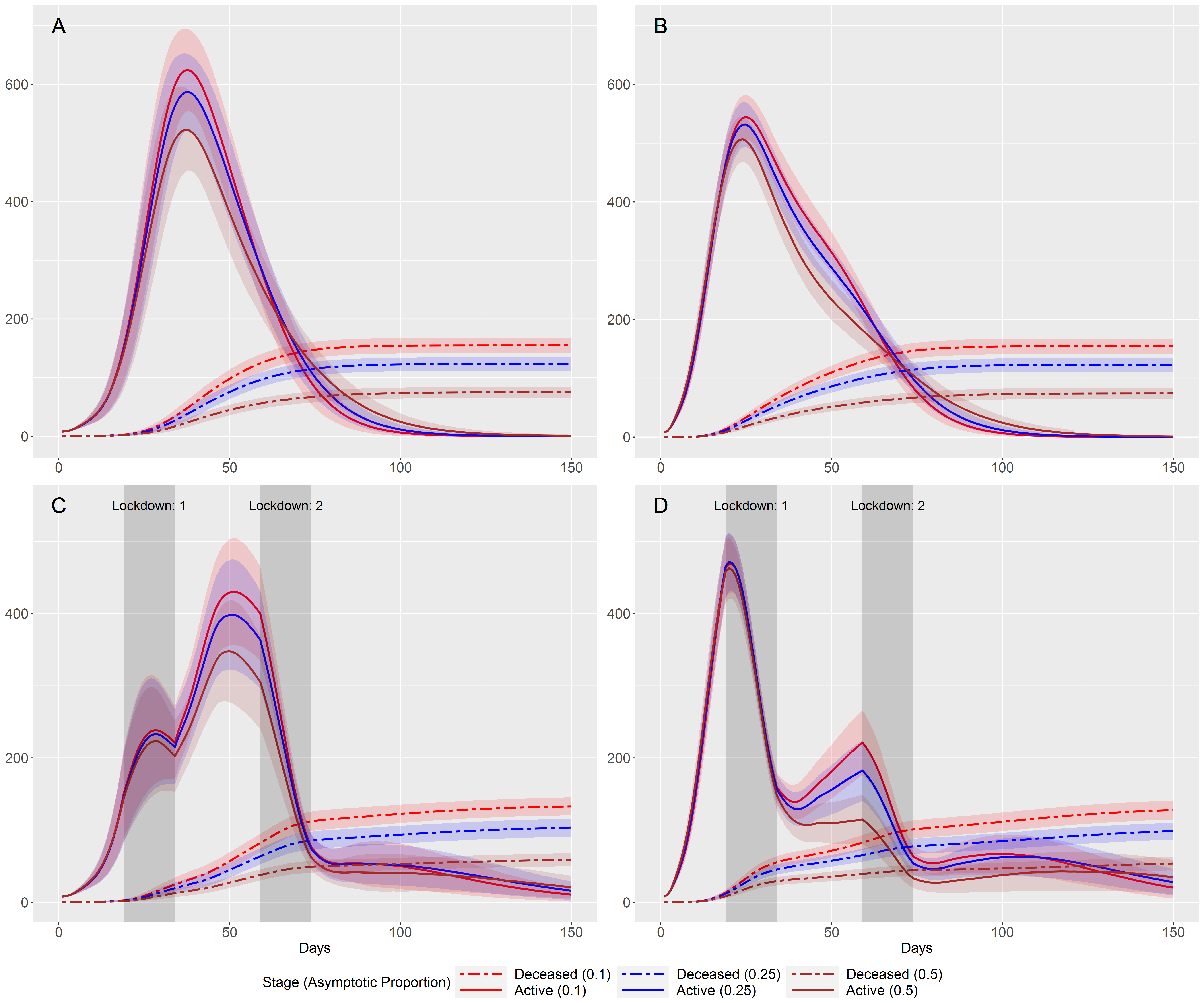}
    \caption{Daily active cases and cumulative deceased cases over time for different asymptotic proportions.}
    \label{fig:cases-2}
\end{figure}

Without any lockdown, when outbreak starts in 2 epicentres outside cluster and chance of detecting infected cases is $95\%$, shapes of curves get altered depending on the ratio of symptomatic and asymptomatic cases (Figure \ref{fig:cases-2}A). As the asymptomatic proportion increases, total number of cases and deaths come down, without any significant lateral shift of the curves. Lesser proportion of asymptomatic cases means steeper slope of curve for cumulative cases. Similar pattern is observed in situation when primary cases are located inside clusters (Figure \ref{fig:cases-2}B).

Once we implement lockdown, curves peak later; overall cases and deaths come down with rise in proportion of asymptomatic cases (Figures \ref{fig:cases-2}C, \ref{fig:cases-2}D). When outbreak starts in population cluster, the course is long-drawn even with lockdown; nevertheless total cases and deaths are fewer if asymptomatic infected persons increase.

In spite of having `double hump' of active cases with two lockdowns (Figure \ref{fig:cases-2}C) total cases and deaths come down with increase in asymptomatic proportion, even when the chance of detection is $95\%$ and both the two primary cases located outside cluster. After 150 days of simulation, number of cumulative recovered cases is around 1500 with $10\%$ asymptomatic and decrease to 1250 with $50\%$ asymptomatic infection; number of deaths also comes down from around 125 to around 30-50 cases. Similar pattern in reduction in cases or deaths is seen if primary cases are located inside cluster (Figure \ref{fig:cases-2}D). However in neither of situations, cumulative recovered cases plateaued after 150 days.

Hence, the model not only tests the effectiveness of an intervention like lockdown \cite{das2020critical}, it also predicts the pattern of outbreak with varying proportion of asymptomatic cases or chance of identifying cases in a specific community. With decreasing sensitivity, number of active cases increase. Even with two lockdowns, failure to detect cases would result in more active cases, as well as number of deaths.

\section{Discussion}
Primary impetus to develop an agent based model was to study the effect of potential interventions on the course of COVID-19 disease in a community. Inter and intra country variation in time, place and person distribution of the disease pointed towards a complex interplay of multifarious determinants related to individual and societal perspectives.

To accommodate variability that exists in the real-life situation, our model considered a community mimicking the Indian diaspora. It is, no doubt, difficult to capture the exact flavor of the extreme diversity of this country. We used the 2011 census distribution of 2011 for age and sex distribution in a community.

Occupation of an individual, her/his mobility in connection with daily activities were thought to influence the risk of contracting or transmitting the disease. Intermixing pattern definitely plays a role in spreading the disease in addition to biological and genetic factors. We factored in individual mobility as a marker of intermixing in this model. In a community this mobility or intermixing is not uniform; it varies with habitation, poverty, gender role, work culture, social customs, availability and access to basic amenities etc. In areas where people need to move to crowded places for basic needs like food, water, and sanitation, they are definitely more at risk, as these activities are practically non-negotiable. Ventilated housing conditions common for slum dwellers may also escalate the risk.

Initially the disease influx was prompted by immigrants returning from endemic countries; this was followed by concentrated foci in urban pockets, more so in slums with high population density. Gradually the disease started to engulf per-urban areas of the country, and rural spread appears obvious and ominous.

While designing the model, flexibility to accommodate relevant individual traits or variables was given utmost importance. A person's health behavior, risk and vulnerability are influenced by a variety of factors. Her/his individual personal traits and also the socio-cultural milieu where s/he belongs are both critical. We thought it intriguing and important to study the role of an agent with a particular mix of personal characteristics in COVID-19 disease spread in her/his community, either as a perpetrator of disease spread or as its victim.

If we consider a local community as a watertight compartment on its own, we can set some rules of interaction within its members and that is exactly what has been done here. Once the primary case of COVID-19 was introduced, the simulation started to run, and from those epicenters, the disease ran its own course based on behavior and personal characteristics of individuals in the community.

Chances of infecting close contacts or getting infected, innate or acquired immunity against the disease, and change in the disease occurrence and mortality over time all can be factored into this model. We thought of building a platform over which simulations with varying individual and group characteristics can be manipulated to observe the effect on the course of the disease over coming months or years. Till date it appears robust enough to allow experimentation till six months after the introduction of the first infected case in a small community. It would be useful to look for a feasible, practical intervention to curb the epidemic spread.

Vaccines generally play an important role in infectious disease epidemiology. However the last pandemic taught us a very important lesson on the need for concerted effort from multiple disciplines. It is critical to run tests during the early phase of the outbreak to identify the exact virus or pathogen, decipher its genetic code and reverse engineer vaccine antigens at the earliest.  The virus may mutate over time to adapt for survival in the community, and the transmission dynamics can also alter. Early detection of target antigen for vaccine and monoclonal antibody development is  the focus of research and innovation in the field of medical countermeasure for pandemics. This model can help to optimize timing of vaccination, benchmark efficacy, and prioritize beneficiary groups.

It can also test for the sustainability of interventions. It is generally accepted that working from home, home quarantine, and physical distancing or lockdown are beneficial to reduce spread. However, after a certain period of time, lockdown appears to become counterproductive with disastrous effects on the livelihood of millions. Moreover, the psychological state of general people resists lockdown over a longer duration. So this model is also useful to study the optimum timing and duration of lockdown on the eventual course of the disease following a peak.

We would like to use this simulation to translate lessons learned into potent measures prior to future outbreaks. The evidence pool generated by basic scientists, laboratories, clinicians, social scientists, demographers and program managers can be incorporated into this robust but flexible simulator. The simulator can definitely help researchers and scientists to test a motley intervention as well as help the health managers and policymakers to decide on optimum timing and duration of effective programmatic measures.

Course of a disease often gets altered with time owing to factors like genetic diversification, modification of host immune response, increased community awareness, changes in environmental parameters or advanced treatment modalities. We assume a pattern of disease spread as a function of all those factors over time.

An agent with assigned characters behaved in a certain way in terms of acquiring or transmitting the infection. In a population with known distribution of those characteristics, this can forecast possible shape of disease outbreaks like COVID-19, which is transmitted from person to person without any intervening vector.

This model forecasts disease spread over time in a specific community with known values for certain parameters. It also provides an idea on the best possible measures to reduce spread and control the outbreak in a community. It can be useful in estimating value of $\text{R}_0$ over time. The model is quite flexible in accommodating 20 different parameters like infectivity period, yield of testing, socio-economic strata, daily travel, awareness level, population density, social distancing, lockdown etc. The same simulation can be used in a synthetic population to study other infectious disease spread. With time, more effective treatment and vaccines will hopefully reduce both the infectivity period as well as the pool of susceptible population. Network analysis for the outbreak in a community might help in further refining the model.

\subsection{Towards AI-enabled learning platforms}
Our simulator already exposes interventions and other external factors as tunable inputs. It returns quantitative outcomes such as active cases, recoveries and deaths. Together, these features provide a natural basis for artificial-intelligence methods. First, the simulator can be cast as a reinforcement-learning environment. Within that framework, the epidemic and demographic configuration represents the state. The choice and timing of interventions (lockdown initiation, duration, and testing intensity) serves as actions. A reward for this learning environment combining averted infections and deaths with the socio-economic cost of restrictions can be optimized to learn adaptive, rather than fixed, intervention policies. Second, the large bank of simulation runs generated through $72$ parameter configurations replicated $1000$ times, provides a labeled dataset on which a machine-learning emulator like a neural network can be trained to approximate outbreak trajectories. Such an emulator could provide faster predictions and support rapid sensitivity analysis and real-time use. Third, the same synthetic-data-generating capability can also support likelihood-free simulation-based inference for calibrating parameters to observed data. Embedding the model within such an AI workflow would help capture the non-linear dynamics of a real outbreak, evaluate the cost-effectiveness of competing measures, and plan multidimensional interventions more effectively across the course of a pandemic. Coupled with continual learning from evolving epidemiological, climatic and pathogen-virulence data, an AI-enabled version of this platform could identify new, personalized or general intervention strategies and thereby strengthen pandemic preparedness across different regions of the world.

\section*{Code and data availability}
The simulator is openly available as an interactive web application at \url{https://anikburman.shinyapps.io/ABM_simulator/}, and the full source code (the \texttt{app.R} Shiny application, demographic input data, and a tutorial video) is hosted at \url{https://github.com/Sayak-999/ABM-Simulator}.

\bibliography{sample}

@article{huang2020epidemic,
	Author = {Huang, Y. and Yang, L. and Dai, H. and Tian, F. and Chen, K.},
	Journal = {Bull World Health Organ},
	Title = {Epidemic situation and forecasting of COVID-19 in and outside China},
	Volume = {10},
	Year = {2020}}

@article{leonardi2020self,
	Author = {Leonardi, M. and Horne, A. W. and Vincent, K. and Sinclair, J. and Sherman, K. A. and Ciccia, D. and Condous, G. and Johnson, N. P. and Armour, M.},
	Journal = {Human Reproduction Open},
	Number = {2},
	Pages = {hoaa028},
	Publisher = {Oxford University Press},
	Title = {Self-management strategies to consider to combat endometriosis symptoms during the COVID-19 pandemic},
	Volume = {2020},
	Year = {2020}}

@article{mwalili2020seir,
	Author = {Mwalili, S. and Kimanthi, M. and Ojiambo, V. and Gathungu, D. and Mbogo, R. W.},
	Journal = {BMC Res Notes},
	Title = {SEIR model for COVID-19 dynamics incorporating the environment and social distancing},
	Year = {2020}}

@article{ghosh2020increased,
	Author = {Ghosh, P. and Basheer, S. and Paul, S. and Chakrabarti, P. and Sarkar, J.},
	Journal = {medRxiv},
	Publisher = {Cold Spring Harbor Laboratory Press},
	Title = {Increased Detection coupled with Social Distancing and Health Capacity Planning Reduce the Burden of COVID-19 Cases and Fatalities: A Proof of Concept Study using a Stochastic Computational Simulation Model},
	Year = {2020}}

@article{yang2020modified,
	Author = {Yang, Z. and Zeng, Z. and Wang, K. and Wong, S. and Liang, W. and Zanin, M. and Liu, P. and Cao, X. and Gao, Z. and Mai, Z. and et al.},
	Journal = {Journal of Thoracic Disease},
	Number = {3},
	Pages = {165},
	Publisher = {AME Publications},
	Title = {Modified SEIR and AI prediction of the epidemics trend of COVID-19 in China under public health interventions},
	Volume = {12},
	Year = {2020}}

@article{das2020critical,
	Author = {Das, S. and Ghosh, P. and Sen, B. and Pyne, S. and Mukhopadhyay, I.},
	Journal = {Statistics and Applications},
	Number = {1},
	Pages = {181-196},
	Title = {Critical community size for COVID-19: a model based approach for strategic lockdown policy},
	Volume = {12},
	Year = {2020}}

@article{roberts2015nine,
	Author = {Roberts, M. and Andreasen, V. and Lloyd, A. and Pellis, L.},
	Journal = {Epidemics},
	Pages = {49--53},
	Publisher = {Elsevier},
	Title = {Nine challenges for deterministic epidemic models},
	Volume = {10},
	Year = {2015}}

@article{lakshmi2020factors,
	Author = {Lakshmi P. S and Suresh, M.},
	Journal = {International Journal of Healthcare Management},
	Number = {2},
	Pages = {89--98},
	Publisher = {Taylor \& Francis},
	Title = {Factors influencing the epidemiological characteristics of pandemic COVID 19: A TISM approach},
	Volume = {13},
	Year = {2020}}

@article{cuevas2020agent,
	Author = {Cuevas, E.},
	Journal = {Computers in Biology and Medicine},
	Pages = {103827},
	Publisher = {Elsevier},
	Title = {An agent-based model to evaluate the COVID-19 transmission risks in facilities},
	Year = {2020}}

@article{hoertel2020stochastic,
	Author = {Hoertel, N. and Blachier, M. and Blanco, C. and Olfson, M. and Massetti, M. and Rico, M. S. and Limosin, F. and Leleu, H.},
	Journal = {Nature Medicine},
	Number = {9},
	Pages = {1417--1421},
	Publisher = {Nature Publishing Group},
	Title = {A stochastic agent-based model of the SARS-CoV-2 epidemic in France},
	Volume = {26},
	Year = {2020}}

@article{epstein2012agent,
	Author = {Epstein, B.},
	Journal = {Models, simulations, and representations},
	Pages = {115--144},
	Publisher = {Routledge},
	Title = {Agent-based modeling and the fallacies of individualism},
	Volume = {9},
	Year = {2012}}

@article{supriya2013abm,
	Author = {Kumar, S. and Grefenstette, J.J. and Galloway, D. and Albert, S.M. and Burke, D.S.},
	Journal = {Am J Public health},
	Title = {Policies to reduce influenza in the workplace: Impact assessments using an agent-based model},
	Volume = {103(8)},
	Year = {2013}}

@article{liu2015,
	Author = {Liu, F. and Enanoria, W.T.A. and Zipprich, J. and Blumberg, S. and Harriman, K. and Ackley, S.F. and Wheaton, W.D. and Allpress, J.L. and Porco, T.C.},
	Journal = {BMC Public Health},
	Title = {The role of vaccination coverage, individual behaviors, and the public health response in the control of measles epidemics: an agent-based simulation for California},
	Volume = {15},
	Year = {2015}}

@article{supriya2015abm,
	Author = {Kumar, S. and Piper, K. and Galloway, D.D. and Hadler, J.L. and Grefenstette, J.J.},
	Journal = {BMC Public health},
	Title = {Is population structure sufficient to generate area-level inequalities in influenza rates? An examination using agent-based models},
	Volume = {15},
	Year = {2015}}

@article{espana2018,
	Author = {Espana, G. and Grefenstette, J. and Perkins, A. and Torres, C. and Carey, A.C. and Diaz, H. and Hoz, F. and Burke, D.S. and Panhuis, W.G.},
	Journal = {Sci Rep},
	Title = {Exploring scenarios of chikungunya mitigation with a data-driven agent-based model of the 2014-2016 outbreak in Colombia},
	Volume = {8},
	Year = {2018}}

@article{guclu2016,
	Author = {Guclu, H. and Kumar, S. and Galloway, D. and Krauland, M. and Sood, R. and Bocour, A. and Hershey, T.B. and Nostrand, E. and Potter, M.},
	Journal = {Disaster Med Public Health Prep},
	Title = {An agent-based model for addressing the impact of a disaster on access to primary care services},
	Volume = {10(3)},
	Year = {2016}}

@article{wong2011automatic,
	Author = {Wong, A. and Scharcanski, J. and Fieguth, P.},
	Journal = {IEEE Transactions on Information Technology in Biomedicine},
	Number = {6},
	Pages = {929--936},
	Publisher = {IEEE},
	Title = {Automatic skin lesion segmentation via iterative stochastic region merging},
	Volume = {15},
	Year = {2011}}

@article{desai2020urban,
	Author = {Desai, D.},
	Journal = {ORF Occasional Paper},
	Title = {Urban Densities and the Covid-19 Pandemic: Upending the Sustainability Myth of Global Megacities},
	Volume = {244},
	Year = {2020}}

@manual{world2020advice,
	Institution = {World Health Organization},
	Journal = {World Health Organization },
	Organization = {World Health Organization},
	Title = {Advice on the use of masks in the context of COVID-19: interim guidance, 5 June 2020},
	Year = {2020}}

@article{potdar,
	Author = {Potdar, V. et al. and ILI-SARI Surveillance Team.},
	Journal = {Front Public Health},
	Title = {Pan-India influenza-like illness (ILI) and Severe acute respiratory infection (SARI) surveillance: epidemiological, clinical and genomic analysis},
    Volume = {11:1218292},
	Year = {2023}}

@article{spencer,
	Author = {Spencer, J. A. et al.},
	Journal = {Journal of theoretical biology},
	Publisher = {National Acad Sciences},
	Title = {Distinguishing viruses responsible for influenza-like illness},
    Volume = {545},
	Year = {2022}}

\end{document}